\documentclass[aps,pre,reprint,twocolumn]{revtex4-1}
\usepackage{epsfig}
\usepackage{color}
\usepackage{hyperref}
\usepackage{amsmath}
\usepackage{amssymb}
\usepackage{placeins}

\definecolor{red}{rgb}{1,0,0}
\definecolor{blue}{rgb}{0,0,1}
\definecolor{black}{rgb}{0,0,0}

\def\fv{\vec{f}}
\def\Fv{\vec{F}}

\def\Vv{\vec{V}}
\def\uv{\vec{u}}
\def\nv{\vec{n}}
\def\Ic{{I}}
\def\Jc{{J}}
\def\Lc{{\cal L}}
\def\Dc{{\cal D}}
\def\Oc{{\cal O}}
\def\Pc{{\cal P}}
\def\edot{\dot\epsilon}

\newcommand{\eq}[1]{\begin{align}#1\end{align}}

\newlength{\arrow}
\newcommand\blfootnote[1]{%
  \begingroup
  \renewcommand\thefootnote{}\footnote{#1}%
  \addtocounter{footnote}{-1}%
  \endgroup
}

\begin{document}
\title{Effect of Friction on Dense Suspension Flows of Hard Particles}

\author{M. Trulsson${}^1{}^\dagger$\blfootnote{${}^\dagger$ These authors contributed equally to this work.}, E. DeGiuli${}^2{}^\dagger$, and M. Wyart${}^2$}
\affiliation{
$^1$Theoretical Chemistry, Department of Chemistry, Lund University, Sweden \\
$^2$Institute of Theoretical Physics, \'Ecole Polytechnique F\'ed\'erale de Lausanne (EPFL), CH-1015 Lausanne, Switzerland
}

\begin{abstract}
We use numerical simulations to study the effect of particle friction on suspension flows of non-Brownian hard particles. By systematically varying the microscopic friction coefficient $\mu_p$ and the viscous number $J$, we build a phase diagram that identifies three regimes of flow: \emph{Frictionless, Frictional Sliding,} and \emph{Rolling}. Using energy balance in flow, we predict relations between kinetic observables, confirmed by numerical simulations. For realistic friction coefficient and small viscous numbers (below $J\sim 10^{-3}$) we show that the dominating dissipative mechanism is sliding of frictional contacts, and we characterize asymptotic behaviors as jamming is approached. Outside this regime, our observations support that flow belongs to the universality class of frictionless particles. We discuss recent experiments in the context of our phase diagram.
\end{abstract}
\maketitle

\section{Introduction}

Non-Brownian dense suspensions are central to industrial processes including oil extraction and food processing,
as well as natural phenomena such as landslides and slurries. Understanding how such out-of-equilibrium, crowded systems flow remains a great challenge, as such systems can show shear-thinning, shear-thickening, and shear-banding depending on both the microscopic nature of the interparticle forces, inertial effects, and the boundary conditions \cite{Fall10,Trulsson12,Picano13,Seto13,Wyart14}. 
To make progress in such complex systems, two limiting cases have received considerable attention. 
On the one hand, for {\it infinitely hard frictional} particles, dimensional analysis alone implies that the pressure carried by the particles, $p$, and shear rate $\edot$ do not affect the flow independently, but only through the viscous number $\Jc = \eta_0 \edot/p$ where $\eta_0$ is the solvent viscosity \cite{Lemaitre09b}. In particular, the packing fraction $\phi$ and the macroscopic friction $\mu = \sigma/p$ must be  functions of $\Jc$, and are found empirically to follow constitutive laws of the type $\phi(\Jc) = \phi_c - a_\phi \Jc^{\gamma_\phi}$ and $\mu = \mu_c+a_\mu \Jc^{\gamma_\mu}$ \cite{Boyer11}. These laws imply a quasi-Newtonian behavior at fixed $\phi$, with a viscosity $\eta/\eta_0 \propto (\phi_c - \phi)^{-\gamma}$ with $\gamma=1/\gamma_\phi$. Currently these relations are phenomenological, with $\gamma=1/\gamma_\phi\approx1/\gamma_\mu\approx 2$.  On the other hand, for {\it frictionless particles} more precise numerical measurements were made \cite{Peyneau09,Olsson11,Olsson12} and report values $\gamma=1/\gamma_\phi\approx1/\gamma_\mu\in [2.5, 2.8]$. A microscopic theory can rationalize these findings and predicts $\gamma=2.85$ and $\gamma_\phi=0.35$ 
\cite{DeGiuli15a,During16}. The central idea is that as the density increases, the network of contacts between particles becomes more and more constrained: there are fewer and fewer floppy modes along which particles can flow without overlapping \cite{Lerner12a}. As the number of floppy modes vanishes, nondimensionalized velocity fluctuations diverge as $\Lc\sim \Jc^{-1/2}$ \cite{Lerner12a} and become correlated on a length scale $\ell_c\sim \Jc^{-0.15}$ \cite{During14,DeGiuli15a}.
These results assume that dissipation stems only from viscous forces; the extent to which they apply to frictional particles is unknown. 

\begin{figure}[th!]
\hspace*{-0.7cm}\includegraphics[width=1.1\columnwidth]{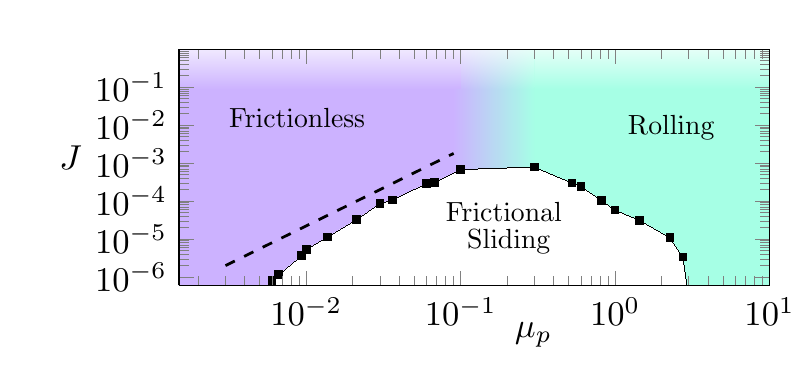}
\caption{Phase diagram of dense non-Brownian suspension flow. In the Frictionless and Rolling regimes, the dominant source of dissipation is viscous drag, whereas in the Frictional Sliding regime, dissipation is dominated by sliding friction. The dashed line has slope 2.}
\label{fig1}
\end{figure}

Similar considerations apply to {\it inertial} flow of hard particles: the rheology has a singular dependence on the dimensionless strain rate, the inertial number $I=\edot D \sqrt{\rho/p}$, where $D$ is particle diameter and $\rho$ is particle density. Empirically $\phi(I) = \phi_c - a_\phi I^{\alpha_\phi}$ and $\mu(I) = \mu_c+a_\mu I^{\alpha_\mu}$, where $\alpha_\mu \approx \alpha_\phi \approx 0.38$ for frictionless particles \cite{Peyneau08}, while $\alpha_\mu \approx \alpha_\phi \approx 0.85$ for frictional particles \cite{Peyneau09}. In this case, a theory can explain the flow behavior in the frictionless regime \cite{DeGiuli15a}, where dissipation is due only to inelastic collisions, but the rheology clearly differs when friction is added. For inertial flow we recently worked out the phase diagram when both friction coefficient $\mu_p$ and inertial number $I$ are varied, quantifying the transition from frictionless to frictional behavior \cite{DeGiuli16}. However, it is not clear how these results translate to suspensions. Even if one can empirically map the viscous to the inertial flow curves at certain friction coefficients and at intermediate shear-rates \cite{Trulsson12}, to extend such a map to arbitary friction coefficients and shear-rates one needs to have a detailed knowledge of scaling laws in all regimes, i.e. correct exponents and the range over which these scalings are valid.

In this work, we characterize the role of friction in suspensions of hard particles using a combination of numerics and scaling arguments. We systematically vary the friction coefficient $\mu_p$ and viscous number $\Jc$, and 
establish a transition from regimes where energy dissipation is dominated by viscous forces  to a Frictional Sliding regime in which sliding friction dominates and physical properties differ, as shown in Fig \ref{fig1}. We establish novel scaling behaviors, distinct from the dry granular case\cite{DeGiuli16}, and a scaling relation for the sliding velocity in this regime, and provide indications that dissipation is inhomogeneous in space, unlike in frictionless systems. 

This regime diagram is in striking similarity with the phase diagram of \textit{inertial} flow \cite{DeGiuli16}, where collisional and sliding dissipation compete. However, a key difference is that for inertial flow, the frictional sliding regime ends at $I \approx 0.1$, where it is known that the dense flow regime ends, but as is visible in Fig.\ref{fig1}, for viscous flow the Frictional Sliding regime ends near $\Jc \approx 10^{-3}$. There is thus a large range of viscous number for which dense flows are essentially frictionless. As we discuss in the Conclusion, this sheds light on previous experiments. 

\section{Numerical Protocol} We use the Discrete Element Method to model flow of granular materials with strictly overdamped dynamics; grains are stiff elastic disks with Coulomb friction, subject to drag forces modelling their interaction with the solvent. When grains overlap at a contact $\alpha$, they experience a harmonic elastic force $\fv_\alpha$ with tangential and normal components $\fv_\alpha{}^{\!T}$ and $f_\alpha^N$; the spring constants have a ratio $k_T=0.5 k_N$. Coulomb friction restricts the elastic forces to satisfy $|\fv_\alpha{}^{\!T}| \leq \mu_p f_\alpha^N$; contacts that saturate this constraint are said to be {\it sliding}, while the rest are said to be {\it rolling}. Long-range hydrodynamic interactions are neglected, as such interactions are believed to be screened  by the dense and disordered solid part of the suspension; viscous interactions between the particles and the viscous fluid are modeled by Stokes drag, both forces and torques. We thereby assume laminar flows, which also seems to be the
dominating case for most Reynolds numbers in dense suspensions \cite{Lashgari14} .The insensitivity of bulk rheology to details of the lubrication interactions and long-range hydrodynamics for sufficiently dense flows was previously established \cite{Peyneau08,Andreotti12,Vagberg14b}.

Systems are initially prepared by sedimenting grains under gravity in an $x-$periodic domain. When this is complete, walls are then constructed from strips of grains near the upper and lower edges of the domain. Shear is imposed by horizontal motion of the wall; we perform our numerics at imposed global shear rate and constant pressure. Grain stiffness is such that relative deformation at contacts is $\Delta = p/k_n \approx 10^{-3}$, within the rigid limit established previously \cite{Trulsson12}. We work in 2D, and accordingly use notation in which torques and angular velocities are scalars.

Since dynamics is strictly overdamped, each particle is subject to equations of force and torque balance. The former is $0 = \Fv^{ext}_i + \Fv^v_i - \sum_{j} \fv_{ij}$, where $\Fv^{ext}$ is the external force, $\Fv^v_i$ is the viscous drag force, and $\fv_{ij}$ is the contact force exerted on particle $j$ by particle $i$. The torque balance equation is $0 = \tau^{ext}_i + \tau^v_i - \sum_{j} \tau_{ij}$. For drag forces, we consider a simple Stokes drag $\Fv^v_i = -\eta_0 c_d (\Vv_i - \Vv^a_i)$ and $\tau^v_i = - \frac{4}{3} \eta_0 c_d R_i^2 (\omega_i - \omega^a_i)$ where $R_i$ is radius of particle $i$, $\Vv^a_i = \edot y {\hat x}$ is the affine velocity, and $\omega^a_i = \edot/2$. Here $c_d=3 \pi/(1-\phi_0)$,  with $\phi_0=0.76$, is a constant chosen to scale the drag contributions $\Jc$ to experiments \cite{Trulsson12}. Its importance is discussed in the Discussion section below.

The Coulomb friction coefficient is varied from 0 to 10. For realistic grains, $\mu_p$ is often reported in the range $0.2$ to $0.7$, however smaller values may be relevant for emulsions, foams, and coated surfaces, and larger values may model the effect of particle angularity \cite{Estrada11}.

\begin{figure}
\includegraphics[width=\columnwidth]{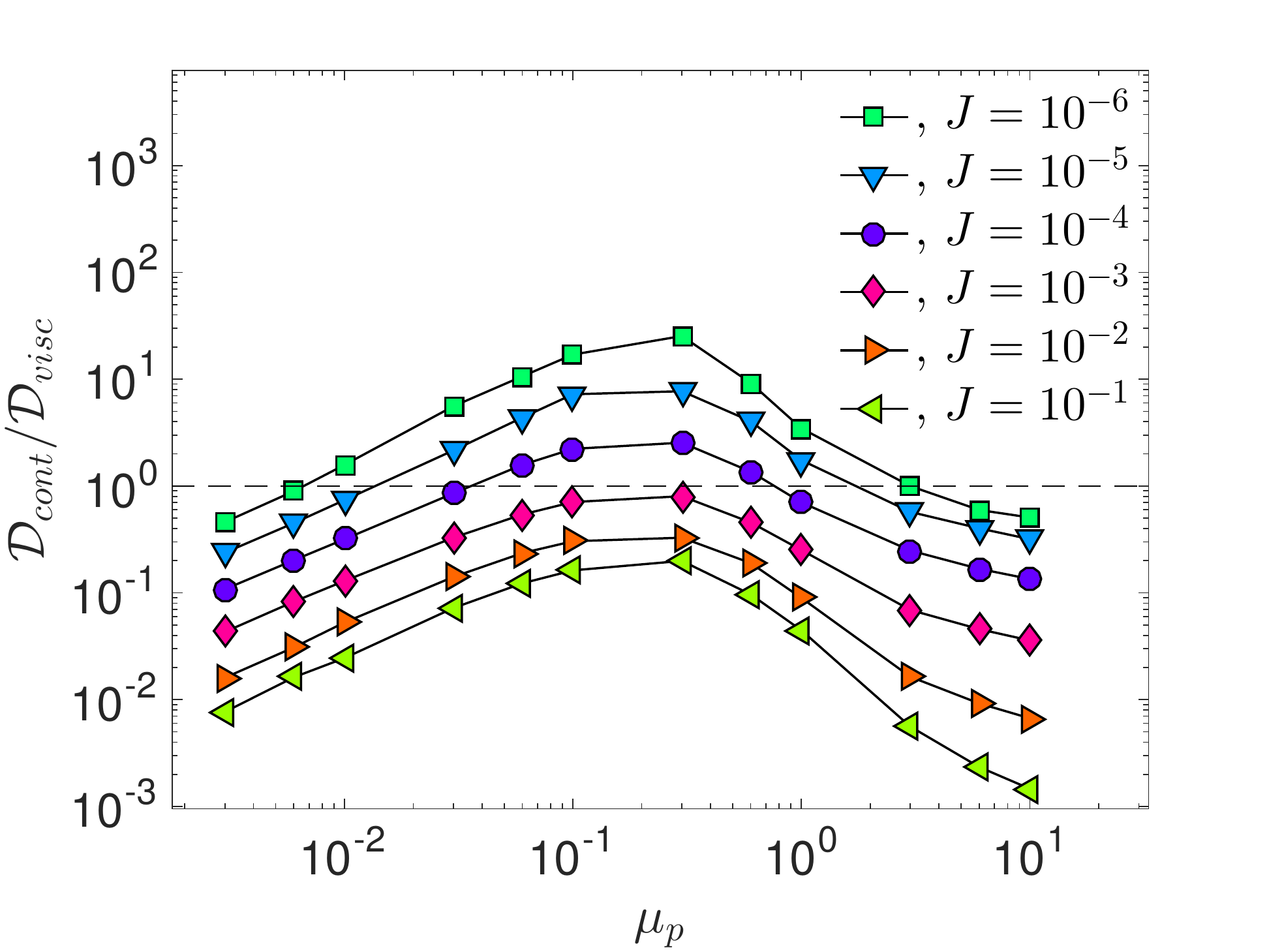}
\caption{Ratio of dissipation induced by sliding at contacts to viscous dissipation at indicated viscous numbers. Unity discriminates between flows that are dominated either by frictional or non-frictional dissipation.}
\label{fig2}
\end{figure}

\begin{figure*}[th!]
\includegraphics[width=\textwidth]{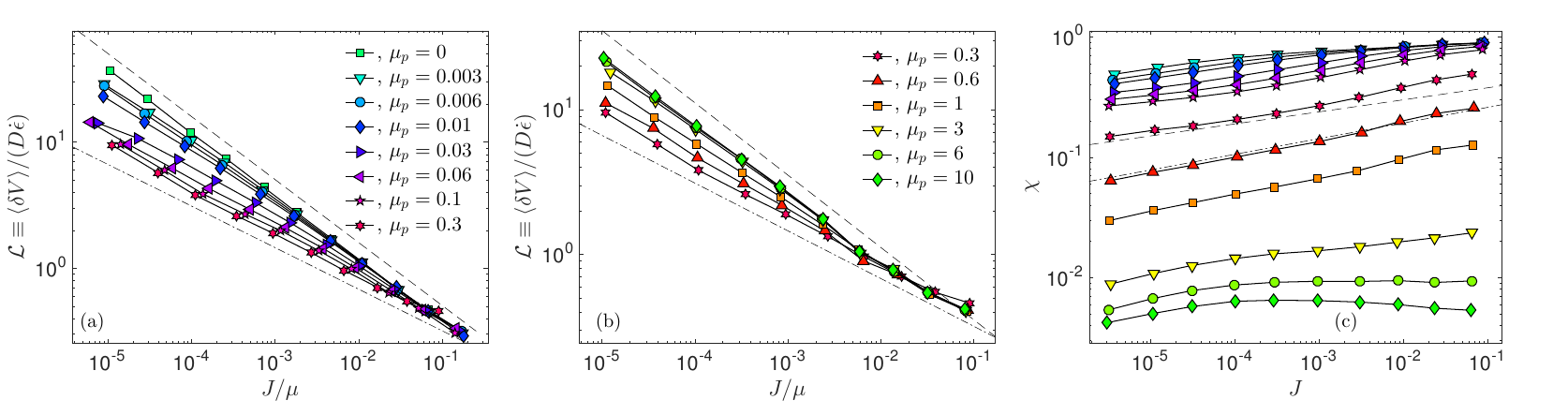}
\caption{(a,b) Relative fluctuations around the affine velocity field, $\Lc$, as function of the viscous number at various particle friction coefficients. Dashed and dot-dashed lines indicate slopes $-1/2$ and $-1/3$, respectively. (c) Fraction of sliding contacts, $\chi$. Dashed and dot-dashed lines indicate slopes $0.1$ and $0.13$, respectively. }
\label{fig3}
\end{figure*}

\section{Energy balance} 
\begin{figure}[bh!]
\includegraphics[width=\columnwidth]{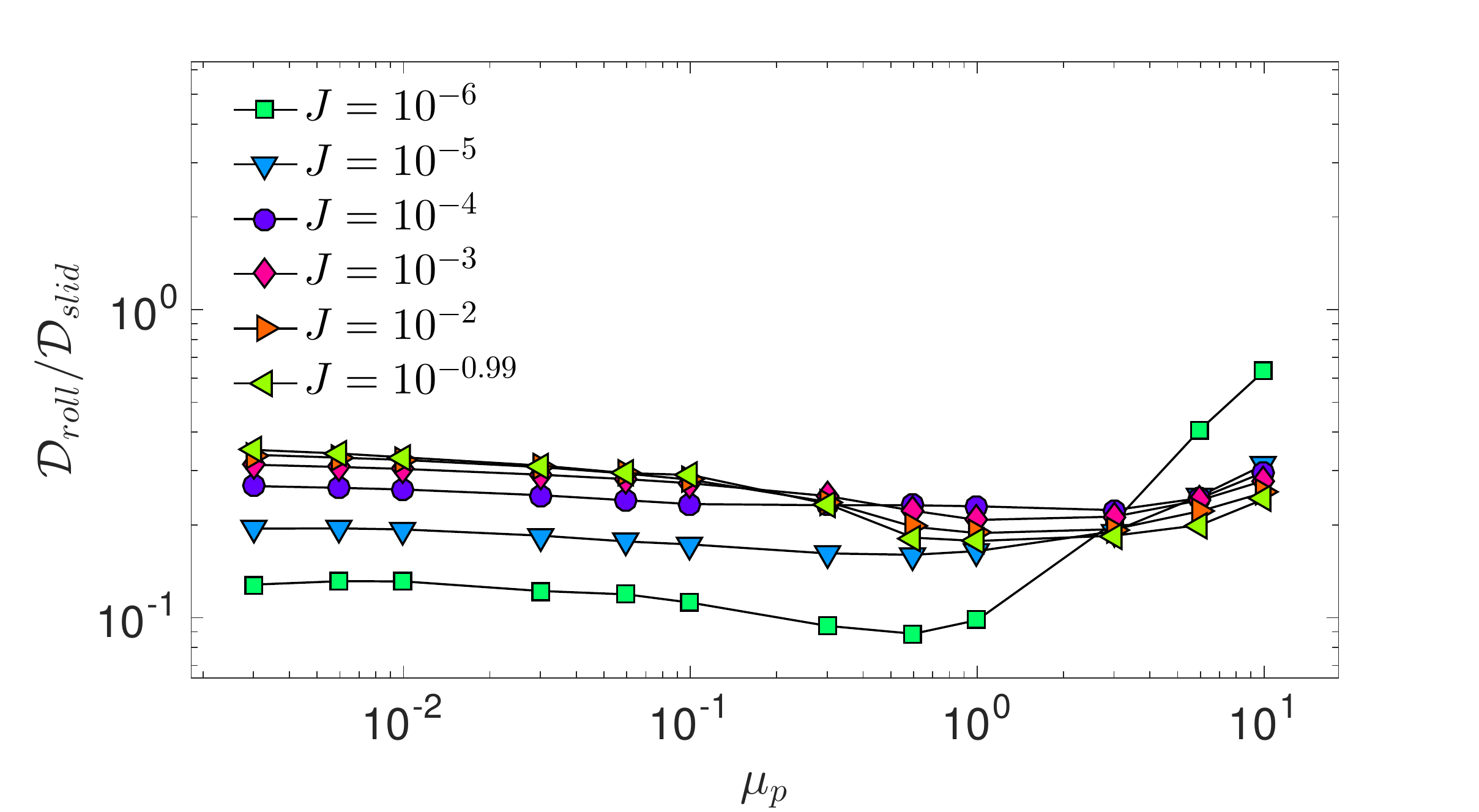}
\caption{Ratio of sliding dissipation at rolling contacts to that at sliding contacts.}
\label{fig4new}
\end{figure}

Energy conservation is a strong constraint on stationary flows, as we now show. The energy balance equation is obtained by contracting the force and torque balance equations along the velocity field $(\Vv_i, \omega_i)$. The result is 
\eq{ \label{power}
\Pc_{ext} = \Dc_{visc} + \Dc_{cont},
}
where $\Pc_{ext} = \sum_i \Fv^{ext}_i \cdot \Vv_i + \tau^{ext}_i \omega_i$ is the power injected through external forces, $\Dc_{visc}$ is the power dissipated through viscosity, and $\Dc_{cont}$ is the net power dissipated through contact forces. The viscous term is
\eq{ \label{Dv}
\Dc_{visc} & = c_d \eta_0 \big[ \sum_i \Vv_i \cdot (\Vv_i - \Vv^a_i) + \frac{4}{3} R_i^2 \omega_i \cdot (\omega_i - \omega^a_i) \big] \notag \\
& = c_d \eta_0 \big[ \sum_i (\Vv_i - \Vv^a_i)^2 + \frac{4}{3} R_i^2 (\omega_i - \omega^a_i)^2\big] , 
}
since non-affine and affine velocities are uncorrelated on average, i.e. $\sum_i \Vv^a_i \cdot (\Vv_i - \Vv^a_i) = 0$, and similarly for the torques. The contact term is
\eq{ \label{Dc}
\Dc_{cont} & = - \sum_i \Vv_i \cdot \sum_{j} \fv_{ij} - \sum_i \omega_i \sum_{j} \tau_{ij} \notag \\
& = \sum_{ij} \uv_{ij} \cdot \fv_{ij},
}
where $\uv_{ij} = \Vv_j - \Vv_i + \nv_{ij}^\perp (R_j\omega_j  + R_i \omega_i )$ is the relative velocity of particle $j$ with respect to particle $i$ at their mutual contact point; $\nv_{ij}^\perp$ is a vector perpendicular to the contact normal $\nv_{ij}$. The second equality in Eq.~\eqref{Dc} can be established by direct substitution of the definition of $\uv_{ij}$. The work done by contact forces has both normal and tangential components. The former is entirely elastic and vanishes on average, reflecting the fact that elastic forces are conservative. 
We are left with:
\eq{ \label{Dc2}
\Dc_{cont} & = \sum_{ij \in C_S} \uv_{ij}^T \cdot \fv_{ij}^T + \sum_{ij \in C_R} \uv_{ij}^T \cdot \fv_{ij}^T \notag \\
& \equiv \Dc_{slid} + \Dc_{roll},
}
where $C_S$ denotes the sliding contacts, $C_R$ denotes the non-sliding (rolling) contacts. The term  $\Dc_{roll}$ can be shown to vanish on average if the normal force in the contact was held constant, as it would correspond to the loading of a transverse spring whose mechanical energy must be bounded. However, it can be finite for contact force history that include sliding periods. We find empirically that for all parameters probed $\Dc_{roll} \lesssim \Dc_{slid}$, as shown in Fig. \ref{fig4new}, and $\Dc_{roll}$ will thus be neglected in scaling arguments below. 

To investigate which source of dissipation dominates, we plot the ratio $\Dc_{cont}/\Dc_{visc}$ in Fig.\ref{fig2}. 
Following our previous work \cite{DeGiuli16}, we can define regime boundaries at the viscous number at which $\Dc_{cont}/\Dc_{visc}=1$. The result is plotted in Figure \ref{fig1}. We find three regimes, as discussed above. Note that the transition from Frictionless to Rolling is a crossover: throughout this transition area, dissipation is dominated by viscous drag, but the structure of the contact network depends strongly on $\mu_p$, as discussed below.

\section{Scaling estimates of dissipation} In steady state, energy input from the shear stress is $\Omega \sigma \edot$, where $\Omega$ is the system volume and $\sigma$ the shear stress. For large systems, additional contributions from fluctuations of the normal position of the wall are insignificant, thus 
\eq{
\Pc_{ext} = \Omega \sigma \edot.
}
To estimate the dissipation rate from viscous dissipation, we consider the velocity scale $\delta V$ defined as the square root of the time-averaged second moment of the non-affine velocity, \textit{i.e.} $\delta V= \langle (\Vv_i-\Vv^a_i)^2\rangle^{1/2}$. We expect that the scale of angular velocity fluctuations is then $\delta \omega \approx \delta V (2/D)$, where $D$ is the mean particle diameter. From Eq.\ref{Dv} we then have $\Dc_{visc}\approx 7c_d \eta_0 N \delta V^2/3$, where $N$ is the number of particles in $\Omega$. It is convenient to define dimensionless dissipations per particle, normalizing by $\Omega p \edot \approx N \pi (D/2)^2 p \edot/\phi$. Thus we let  $\tilde{\Dc}_{visc} \equiv \Dc_{visc} / (\Omega p \edot)$, so that
\eq{ \label{Dv2}
\tilde{\Dc}_{visc} \approx \frac{28 \phi}{3 \pi} \frac{c_d \eta_0 N \delta V^2}{D^2 p \edot} \approx \frac{28 \phi}{3 \pi} c_d \;  \Lc^2 \Jc,
}
where we defined $\Lc \equiv \delta V /(\edot D)$. Similarly, the sliding dissipation rate can be estimated from Eq.\ref{Dc} as $\Dc_{slid} \approx N_c \chi u_T \mu_p f_N$, where $N_c$ is the number of contacts, $\chi$ is the fraction of sliding contacts, $u_T$ is the mean sliding velocity, and $f_N$ is the mean normal force. The pressure is related to the normal force by $p \approx N_c f_N D/(2\Omega)$. Defining $\tilde{\Dc}_{slid} \equiv \Dc_{slid} / (\Omega p \edot)$ we find
\eq{ \label{Dslid}
\tilde{\Dc}_{slid} \approx \frac{\chi \mu_p 2 p \Omega u_T/D}{\Omega p \edot} = 2 \mu_p \chi \Lc_T
}
where $\Lc_T \equiv u_T /(\edot D)$. Since $\sigma = \mu p$, Eq.\ref{power} reads $\mu = \tilde{\Dc}_{visc} + \tilde{\Dc}_{cont}$. Using $\Dc_{cont}\sim\Dc_{slid}$ and Eqs.(\ref{power},\ref{Dv2},\ref{Dslid}) yields
\eq{
\mu & \sim C_d \Jc \Lc^2 \qquad & \mbox{Frictionless, Rolling} \label{energy1} \\
\mu & \sim 2 \mu_p \chi \Lc_T, \qquad & \mbox{Frictional Sliding} \label{energy2},
}
with $C_d =28 c_d \phi_c/(3 \pi) \approx 10^2$. We neglect here the variation of $\phi$ with $\Jc$ that induces small corrections to scaling. We now measure the microscopic quantities $\Lc$, $\Lc_T$, and $\chi$ and show that they abide by Eqs.(\ref{energy1},\ref{energy2}).

\begin{figure}[tb!]
\includegraphics[width=\columnwidth]{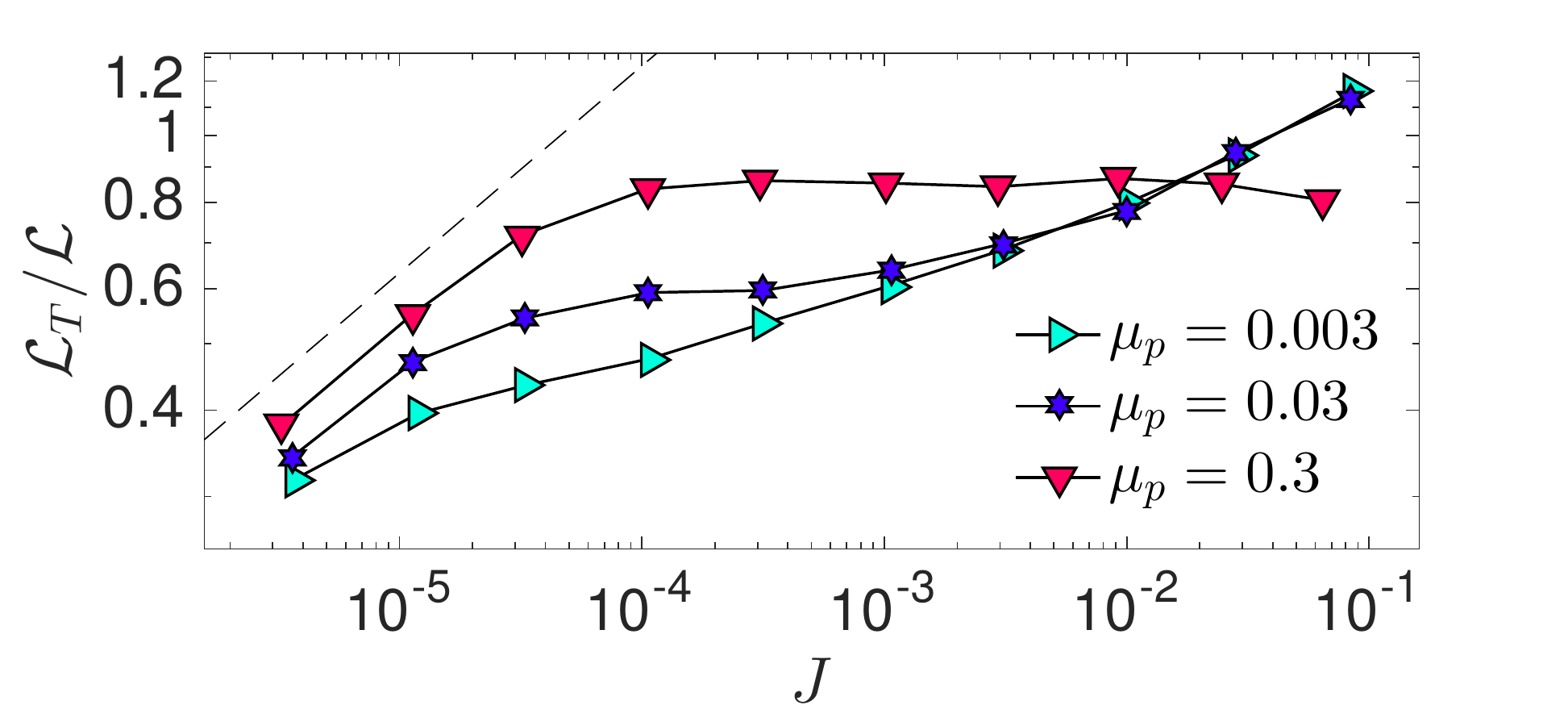}
\caption{ Ratio of mean sliding velocity to velocity fluctuations. Dashed line indicates $\Lc_T/\Lc \sim \Jc^{0.3}$, suggesting flow inhomogeneity once one enters in the sliding frictional regime. }
\label{fig4}
\end{figure}

\begin{figure*}[th!]
\includegraphics[width=\textwidth]{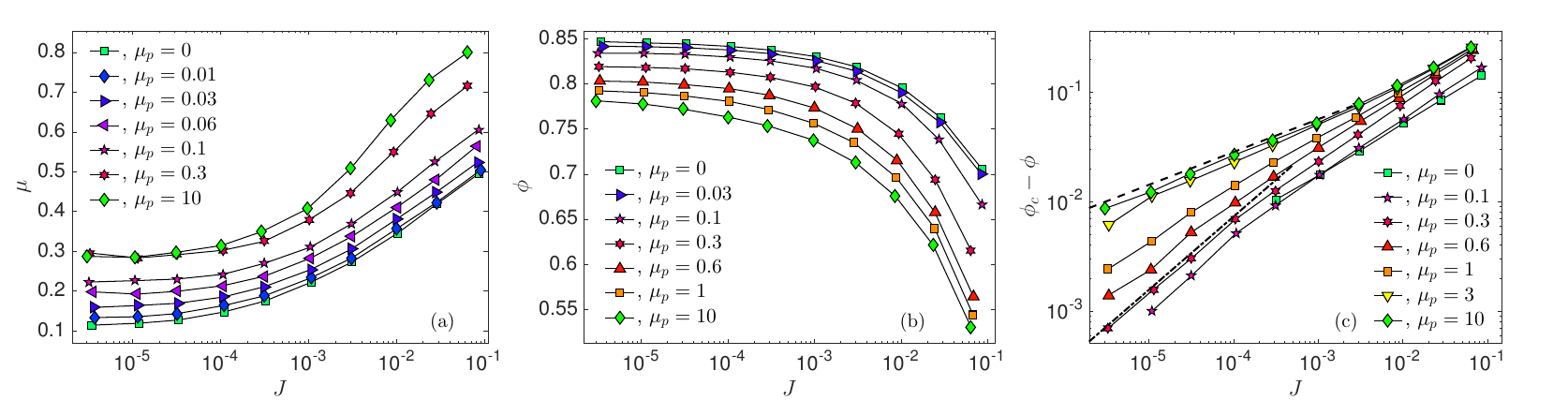}
\caption{(a) Stress ratio $\mu$, (b) volume fraction $\phi$, and (c) $\phi_c-\phi$ as function of the viscous number $\Jc$ at indicated particle friction coefficients. In (c), values of $\phi_c$ were determined by fitting curves in (b). In (c), dashed slope is $0.30$, and dash-dotted slope is $2/3$. 
 }
\label{fig5}
\end{figure*}

\begin{figure}[b!]
\includegraphics[width=\columnwidth]{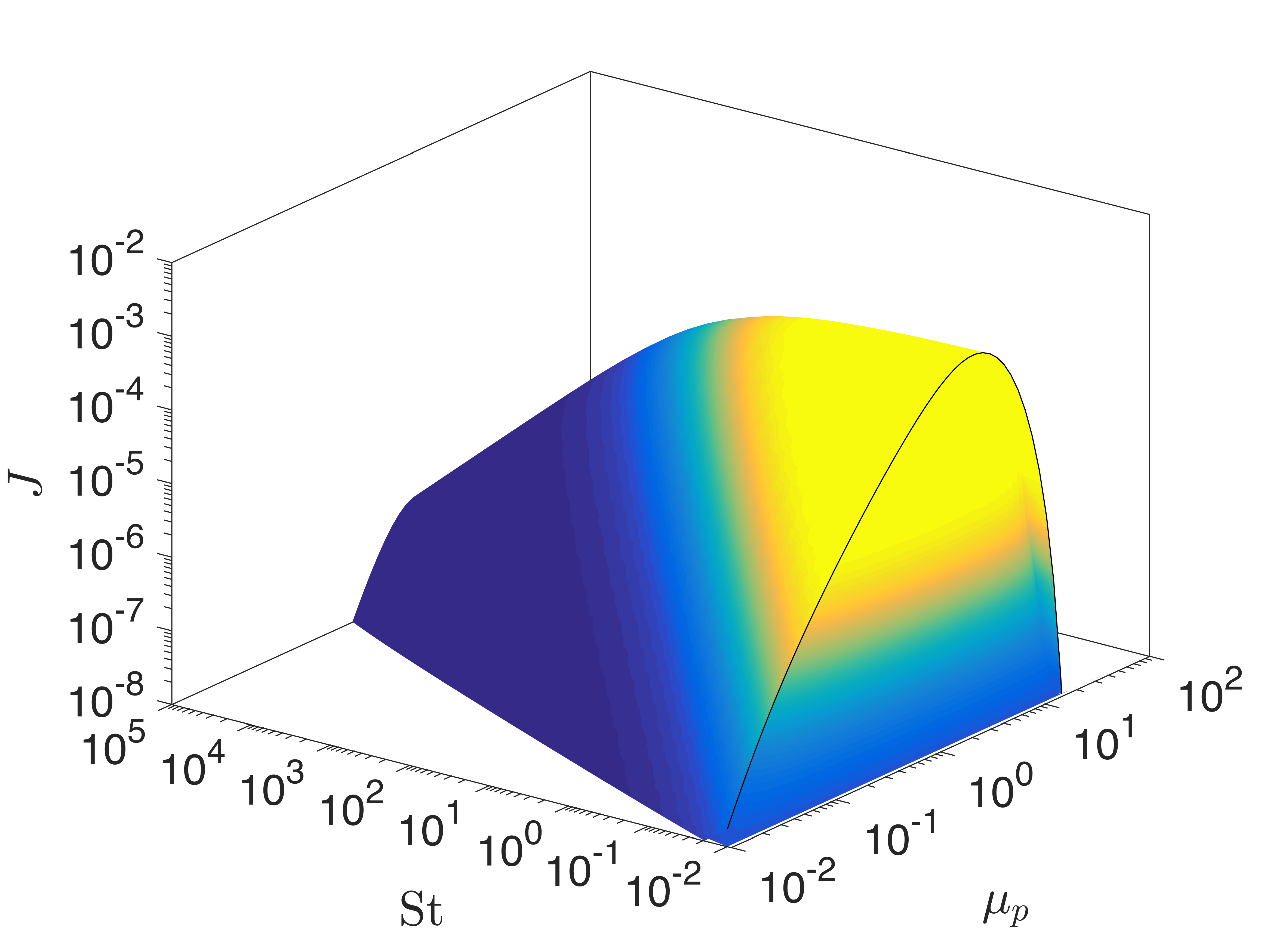}
\caption{ Sketch of phase diagram when inertia is present, in terms of $\Jc$, $\mu_p$, and Stokes number $St = \Ic^2/\Jc$, where $\Ic=\edot D \sqrt{\rho/p}$ is the inertial number. Below the colored dome, dissipation is dominated by frictional sliding, while above it is dominated either by viscous dissipation or grain inelasticity. Color corresponds to the value of rescaled confining pressure $\tilde p = (\Ic/(C_d \Jc))^2$ along the critical surface, distinguishing viscous regimes (yellow, $\tilde p \ll 1$) from inertial regimes (dark blue, $\tilde p \gg 1$). }
\label{figphase3d}
\end{figure}
\begin{figure}[b!]
\includegraphics[width=\columnwidth]{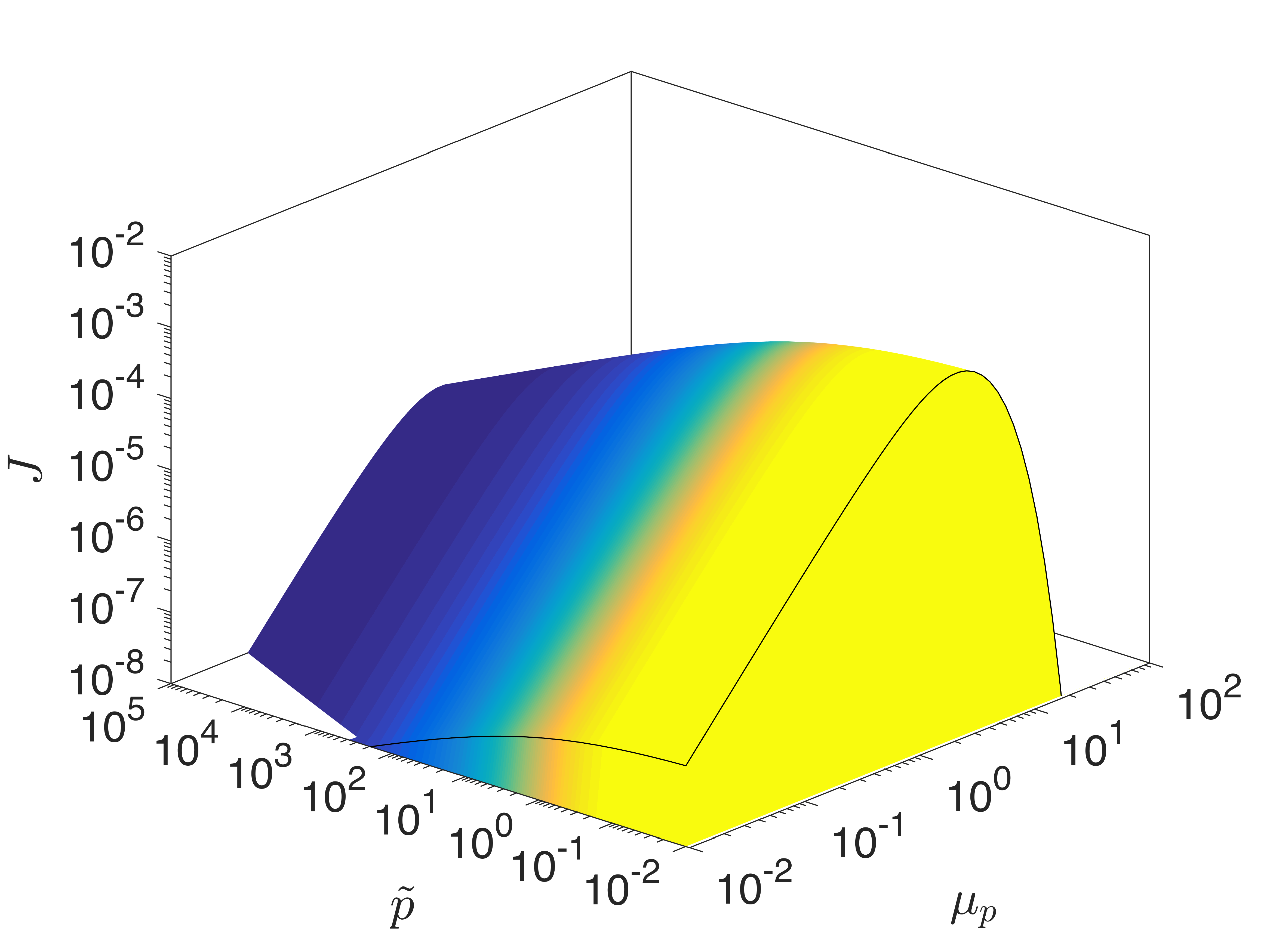}
\caption{ Sketch of phase diagram when inertia is present, in terms of $\Jc$, $\mu_p$, and rescaled confining pressure $\tilde p = p \rho(D/(C_D\eta_0))^2$. Below the colored dome, dissipation is dominated by frictional sliding, while above it is dominated either by viscous dissipation or grain inelasticity. Color corresponds to the value of $\tilde p$ along the critical surface, distinguishing viscous regimes (yellow, $\tilde p \ll 1$) from inertial regimes (dark blue, $\tilde p \gg 1$). }
\label{figphase3dp}
\end{figure}

\section{Microscopic Observables}The quantity $\Lc \equiv \delta V /(\edot D)$ is the ratio of typical velocity fluctuations to the affine velocity scale, shown in Fig. \ref{fig3}. We find that for any $\mu_p$, $\Lc$ increases and apparently diverges as $\Jc$ decreases. In the frictionless and rolling regimes, $\Lc \propto (\Jc/\mu)^{-1/2}$, as follows from Eq.\eqref{energy1}. In the frictional sliding regime, the divergence is weaker: we have approximately $\Lc \sim (\Jc/\mu)^{-1/3}$. 

The typical sliding velocity $u_T$ can differ from $\delta V$ if flow is inhomogeneous. We find evidence for this in the frictional sliding regime, for which a power-law appears to be developing in the accessible range, with approximately $\Lc_T/\Lc \sim \Jc^{0.3}$, as shown in Fig.\ref{fig4}. 

In the frictional sliding regime, an important variable is the fraction of sliding contacts, $\chi$. We plot it in Fig. \ref{fig3}c and show that in this regime it displays a weak scaling: $\chi \sim \Jc^{0.1}$ for $\mu_p=0.3$ and $\chi \sim \Jc^{0.13}$ for $\mu_p=0.6$. Together with the behavior of $\Lc$ and $\Lc_T/\Lc$, this is consistent with the constraint from energy balance: we have $\chi \Lc_T/\mu = \chi (\Lc_T/\Lc) \Lc \mu^{-1} \sim \Jc^{0.1+0.3-0.33} \mu^{0.33-1} = \Jc^{0.07} \mu^{-0.66}$. Since $\mu \approx \mu_c$ in this regime, this is very close to a constant, as predicted by Eq.\ref{energy2}.

In the frictionless and rolling regimes, $\chi$ depends strongly on $\mu_p$. For $\mu_p \gtrsim  1$ we have $\chi < 0.03$, indicating that most contacts are rolling, and justifying the regime's name. We thus find that when crossing from the frictionless to rolling regime, although the velocity fluctuations do not display a strong signal, and the rheology is similar as shown below, the contact network is restructuring from having all sliding contacts to having only rolling contacts.

\section{Rheology}
We now turn to the constitutive relations for the volume fraction of solid material, $\phi(\Jc)$, and the stress ratio $\mu(\Jc)$, shown in Fig.\ref{fig5}. As expected, there is a large dependence on $\mu_p$ in the values of $\mu_c$ and $\phi_c$: the stress ratio increases and the compaction decreases as $\mu_p$ increases at constant $J$. To distinguish flow regimes, we consider in Fig.\ref{fig5}c the quantity $\phi_c-\phi\sim \Jc^{\gamma_\phi}$. In the Rolling regime, we find $\gamma_\phi \approx 0.30$, close to the prediction of \cite{DeGiuli15a,During16} for flows dominated by viscosity. When $\mu_p$ is lowered, but at large enough $\Jc$ to be in the transition from Rolling to Frictionless, we observe that the curves bend.
%
A direct fit for $\mu_p \to 0$ would suggest $\gamma_\phi \approx 0.5$. However, previous analyses in the frictionless limit showed that corrections to scaling are large for frictionless particles, an effect that can be traced back to the smallness of the constant $\mu_c$. If these corrections are taken into account, one finds asymptotic exponents $\gamma_\phi \in (0.36, 0.39)$ \cite{Olsson11,Olsson12,Kawasaki15}. 
We thus suggest that within flow regimes dominated by viscosity, we have  $\gamma_\phi \in (0.30, 0.39)$. In contrast, for intermediate $\mu_p$ the curves display a kink close to the transition from viscous to frictional dissipation; this allows us to fit a much larger slope  $\gamma_\phi \approx 2/3$ within the Frictional Sliding regime.

\begin{figure}[bh!]
\includegraphics[width=\columnwidth]{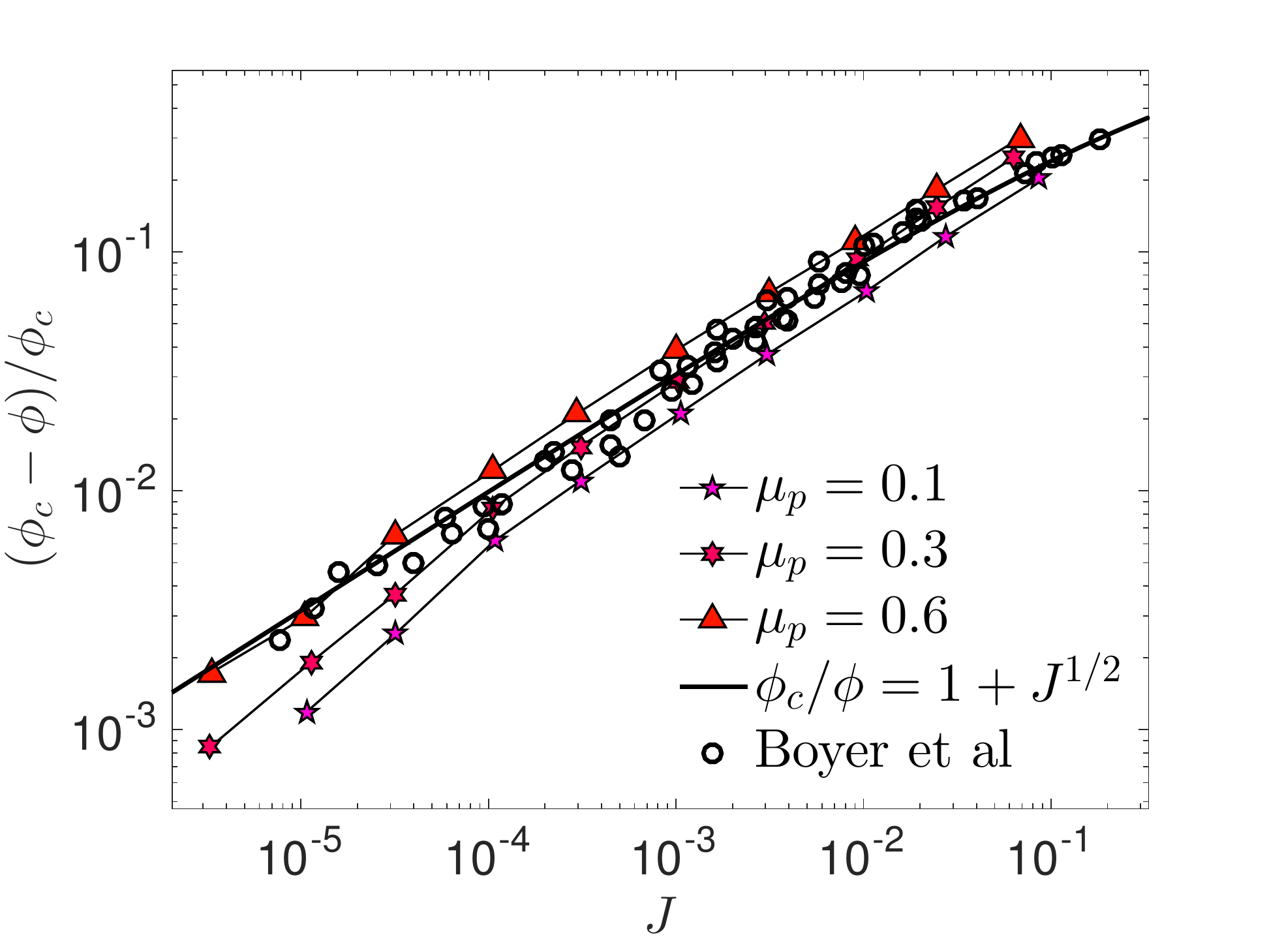}
\caption{ $(\phi_c-\phi)/\phi_c$ as function of the viscous number $\Jc$ at indicated particle friction coefficients, in comparison with data from Boyer et al. \cite{Boyer11}. The solid line shows $\phi_c/\phi = 1 + \Jc^{1/2}$, the fitting form proposed in \cite{Boyer11}. }
\label{figboyer}
\end{figure}

\section{Phase boundaries} We have shown that for realistic $\mu_p$, sliding friction eventually dominates dissipation as jamming is approached. Consistent with the change in dissipation mechanism, key observables differ in the frictional sliding regime: the velocity fluctuations show a weaker divergence with $\Jc$, and the fraction of sliding contacts displays scaling behavior. When $\mu_p \lesssim 0.1$, the crossover to frictional sliding can be understood theoretically: inside the frictionless regime, $\Lc_T \sim \Lc \sim (C_d \Jc/\mu)^{-1/2}$ and $\chi \approx 1$, so that $\tilde{\Dc}_{slid} \approx 2 \mu_p (C_d \Jc/\mu)^{-1/2}$. A crossover to the frictional sliding regime will occur when this quantity is $\Oc(\mu)$. Since $\mu \approx \mu_c$ at small $\Jc$, we see that $\tilde{\Dc}_{slid} \sim \mu_c$ at $\Jc_c \sim 4 \mu_p^2/(C_d \mu_c)$. The scaling of $\Jc_c$ with $\mu_p$, and the small prefactor, are both verified in Fig. \ref{fig1}.

\section{Transition to Inertial flow} It is possible to combine the phase diagram of Fig. \ref{fig1} with earlier results for strictly inertial flow \cite{DeGiuli16} to estimate the transition from viscous to inertial rheology. In inertial dynamics, controlled by the inertial number $I=\edot D \sqrt{\rho/p}$, in addition to viscous dissipation and sliding friction, energy can be dissipated by grain inelasticity. The collisional dissipation rate, which we denote by $\Dc_{coll}$, was previously estimated for hard particles in \cite{DeGiuli15a,DeGiuli16}. Defining, as above, $\tilde \Dc_{coll} = \Dc_{coll}/(\Omega p \edot)$, and considering $\Oc(1)$ restitution coefficient, it reads \cite{DeGiuli15a,DeGiuli16}
\eq{
\tilde \Dc_{coll} \approx \Ic \Lc^2.
}
We assume that this contribution will add to the viscous and frictional contributions, thus modifying power balance from Eq.\ref{power} to
\eq{ \label{powtotal}
\mu \approx \left( C_d \Jc + \Ic \right) \Lc^2 + 2 \mu_p \chi \Lc_T
}
Using the scaling behavior of $\Lc$, $\chi$, and $\Lc_T$ in the various regimes, one can determine from this relation all the possible transitions. 

Consider, for example, the transition when inertia starts to dominate over viscous dissipation  \cite{Trulsson12,Fall10}. From Eq.\ref{powtotal} this will occur when $\Ic > C_d \Jc$. This predicts that the relevant dimensionless number is
\eq{
\tilde p = \left(\frac{\Ic}{C_d \Jc}\right)^2 = p \frac{\rho D^2}{C_D^2 \eta_0^2},
}
which is nominally {\it independent} of strain rate, depending only on the pressure $p$, and fixed parameters. This is consistent with experimental measurements in \cite{Fall10}, where the critical shear stress was found to change by less than a factor of 2 when the strain rate varies over a decade. This gives strong support to the microscopic expressions for $\Dc_{visc}$ and $\Dc_{coll}$.

On the relevance of friction, consider flow inside the frictional sliding regime, where dissipation is dominated by $\Dc_{slid}$. Whether the rheology is viscous, with $\sigma \propto \edot$, or inertial, with $\sigma \propto \edot^2$, depends on the magnitude of the two subdominant dissipation rates, $\Dc_{visc}$ and $\Dc_{coll}$. From Eq.\ref{powtotal} we see that viscous dissipation will be larger provided $\Jc > \Ic/C_d$, which defines a criterion for irrelevance of inertial effects.

Finally, we can sketch the phase diagram when inertial effects are present, in particular the boundary between regions dominated by frictional sliding, and by other sources of dissipation. This boundary defines a function $\Jc = f(\mu_p)$, which from Fig. \ref{fig1} has an approximately inverted parabolic shape on logarithmic axes. It satisfies, in particular, $f(\mu_p) \sim \mu_p^2$ for $\mu_p \ll 1$. If regime change is controlled by dissipation, as we have argued here and in \cite{DeGiuli16}, then in the presence of inertia we need only replace $\Jc$ by $\Jc + \Ic/C_d$. In particular, the frictional/frictionless regime boundary will be modified to $\Jc + \Ic/C_d = f(\mu_p)$. We can check this relation by comparing with the corresponding boundary of the frictional sliding found in \cite{DeGiuli16}, where there is a function $\Ic = g(\mu_p)$. The implied relation $g(\mu_p)=C_d f(\mu_p)$ is consistent with the data for $\mu_p < 1$; for example in \cite{DeGiuli16} it was found that $\Ic \sim \mu_p^2$ at small $\mu_p$, consistent with $f(x) \sim x^2$. Also, the peak value of $\Ic$ having a frictional regime is $\Ic \approx 0.1$, which implies a peak value of $\Jc$ having a frictional regime at $\Jc \approx 0.1/C_d \approx 10^{-3}$, consistent with Fig. \ref{fig1}. 

At fixed Stokes number St$ = \Ic^2/\Jc$, the criterion becomes $f(\mu_p) = \Jc + \sqrt{\Jc \mbox{St}}/C_d$, which can be solved for $\Jc$. The resulting phase diagram is sketched in Fig. \ref{figphase3d} for the range $10^{-2} < $ St $< 10^5$. Below the colored dome, dissipation is dominated by frictional sliding. The color corresponds to the value of $\tilde p$: the yellow regions are viscous, and the blue regions are inertial. 

In fact, the above analysis suggests that the phase diagram could more simply be plotted in terms of $\tilde p$ directly; this is done in Fig. \ref{figphase3dp}. In this representation, the boundary between viscous and inertial regimes is a cut along fixed $\tilde p \approx 1$.

\section{Comparison with experiments} To quantitatively compare our findings with experiments, we need to ensure that our definition of $\Jc$ and the associated normalization of drag forces is reasonably accurate.  We have used a Stokes drag model, where the constant of proportionality is known, analytically, for a {\it force exerted by the fluid in the absence of other particles}. When particles are very dense, the drag force is renormalized by an Archimedes force: the fluid drag is enhanced because most of the volume is occupied by particles themselves, unlike in the case of \emph{one} sphere in a \emph{infinite} volume of fluid \cite{Jackson00,Trulsson12}. This leads to the large constant $C_d \sim 10^2$, which translates to the small transition value $J_c$. By the definition $\Jc = \eta_0 \edot/p$, a change in normalization would correspond to a multiplicative shift in $\Jc$. To verify that our normalization is appropriate, and to compare 2D to 3D, it is natural to measure distance from the transition, i.e., $\Jc$, by comparing constant values of $(\phi_c-\phi)/\phi_c$. To test this procedure, in Fig.\ref{figboyer} we reproduce the data of Boyer et al \cite{Boyer11}, along with the fitting form proposed therein, $\phi_c/\phi = 1+ \Jc^{1/2}$. For $\Jc \gtrsim 10^{-3}$, where most data are taken, the data are close to ours for $\mu_p\in (0.1, 0.6)$. The experiment used spheres of two materials: polymethly methacrylate (PMMA) for $\Jc \in (10^{-5}, 10^{-2})$ and polystyrene for $\Jc \in (10^{-4}, 10^{-1})$, both expected to have a friction coefficient $\Oc(1)$ in dry conditions. Although we cannot strongly discriminate between different $\mu_p$ in this comparison, we conclude that the normalization of $\Jc$ is appropriate.


Our results thus support that experiments probing $\Jc\geq 10^{-3}$ are not in the sliding regime, and that dissipation is dominated by viscous drag. In recent works \cite{Chateau08,Dagois15}, a dimensionless local shear rate, equivalent to $\Lc$, was inferred by considering the rheological properties of suspensions immersed in non-Newtonian fluids.  The analysis of \cite{Dagois15} supports that $\Lc\sim \Jc^{-1/2}$ in the range $10^{-3} < \Jc < 10^{-1}$, as predicted from our phase diagram. 


A few experiments, however, investigate dense flows closer to jamming. In \cite{Boyer11}, mentioned above, some data for $\phi_c-\phi$ extend down to $\Jc \approx 10^{-5}$. These data are close to our simulation results for $\mu_p \in (0.3, 0.6)$. Assuming a $\mu_p$ in this range, our phase diagram indicates that the experiment spans both frictional and viscous regimes. We expect a crossover from $\gamma_\phi \sim 0.35$ (but affected by corrections to scaling, as discussed above) to $\gamma_\phi \sim 0.67$. Since fits to $\phi_c-\phi$ are marred by imprecision in the value of $\phi_c$, especially at small $\Jc$, it would be interesting to focus experimentally on the range $\Jc < 10^{-3}$ and extract $\Lc$ in particular, to seek clear deviations from scaling laws associated with frictionless behavior.

\section{Discussion: rigid clusters and elastoplasticity in granular flows?} In the frictional sliding regime, the mechanism of flow leading to $\Lc \sim \Jc^{-1/3}$ and $\Lc_T/\Lc \sim \Jc^{0.3}$ calls for an explanation, because it implies strong heterogeneity. In the frictionless regime, such heterogeneity is absent, consistent with theory that considers a single velocity scale \cite{DeGiuli15a}. One possibility is that frictional systems become  overconstrained, and flow is {\it elastoplastic}, as in models of amorphous solids \cite{Lin14}. In this scenario, flow can be considered as a series of avalanches. In 2D, such avalanches are similar to slip lines of extension $\xi$ that locally accumulate a strain of order unity. Such avalanches lead to approximately rigid-body motion to an extent $\sim \xi$ perpendicular to the line of slip, so that the non-affine velocity during an avalanche is of order $V_{na} \sim \edot D \xi$, i.e. $\Lc \sim \xi$.  However, within the rigid blocks, the pairwise relative velocity of particles is negligible; thus the typical relative velocity $V_r$ obtains its value only from the slip lines, $V_r \sim V_{na}/\xi$. In this regime we therefore expect $\Lc_R/\Lc \sim 1/\Lc$, where $\Lc_R = V_r/(\edot D)$.

\begin{figure}[t!]
\includegraphics[width=\columnwidth]{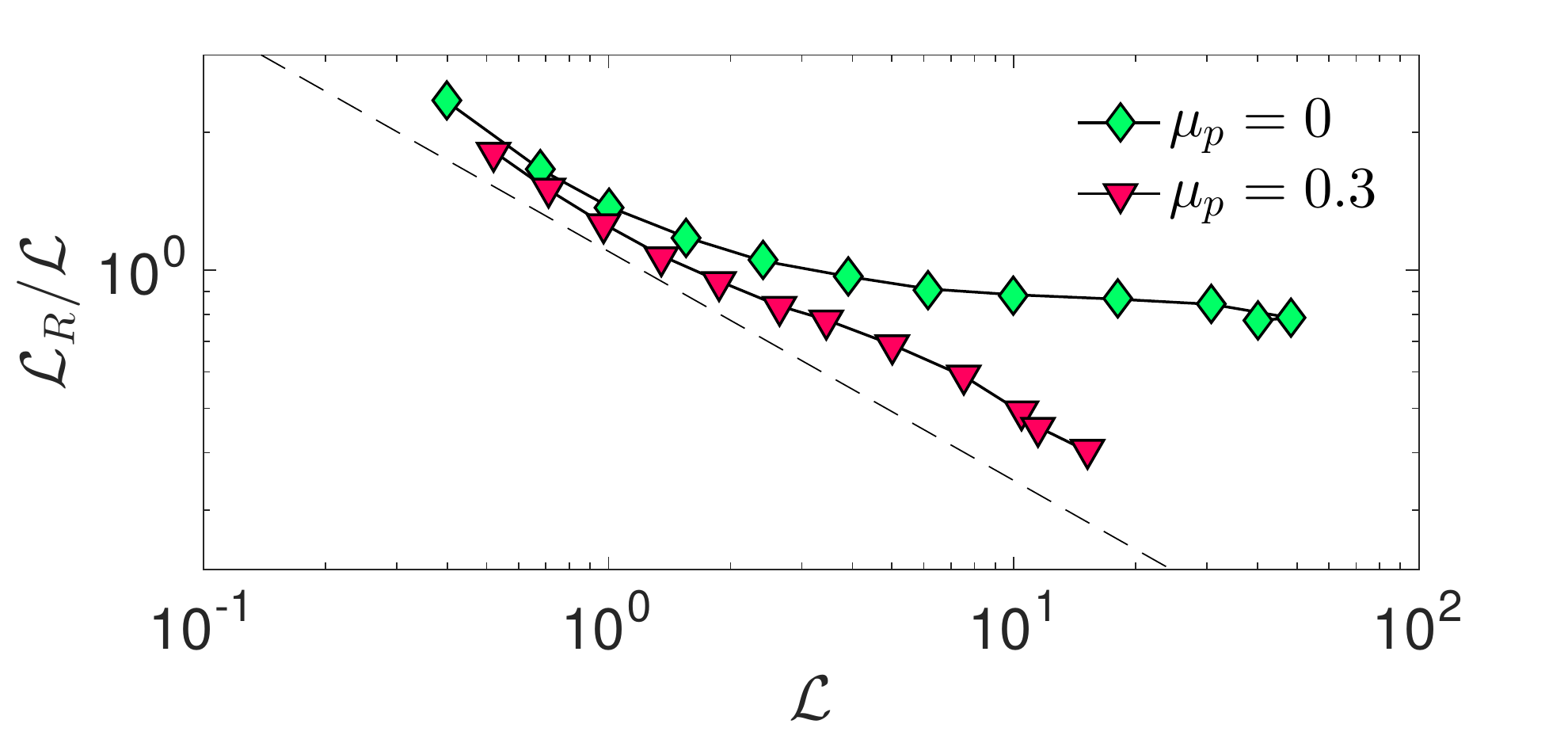}
\caption{ Ratio of typical relative velocity to velocity fluctuations. Dashed line $\Lc_R/\Lc \sim \Lc^{-0.5}$ is a guide to the eye.}
\label{fig6}
\end{figure}

In support of this picture, approximately rigid-body motion has been suggested to occur in granular flows \cite{Heussinger13,Henkes16,Maiti16,Kharel16}. However, scaling relations from elastoplasticity concerning microscopic flow heterogeneity have not previously been tested. Defining $V_r \equiv \langle (V_j-V_i)^2 \rangle^{1/2}$ to be consistent with the definition of velocity fluctuations, we plot  $\Lc_R/\Lc$ for $\mu_p=0$ and $\mu_p = 0.3$ in Fig.\ref{fig6}. When $\Lc$ is small, corresponding to large $\Jc$, particles move nearly affinely, but still nearby particles will have a relative velocity; hence we expect $\Lc_R/\Lc > 1$, as we observe both for $\mu_p=0$ and $\mu_p = 0.3$. However, as $\Lc$ increases from unity, corresponding to denser flow, the curve for $\mu_p=0$ asymptotes near unity, consistent with the theory of \cite{DeGiuli15a}, while the curve for $\mu_p$ continues to decrease, approximately as $\sim 1/\sqrt{\Lc}$ (dotted line). This provides further support for flow inhomogeneity in frictional flow, and moreover appears distinct from the naive elastoplastic prediction  $\sim 1/\Lc$. Our observations thus do not suggest that a simple rigid block picture applies. Further investigations are clearly needed to clarify this point, central to a microscopic understanding of flow.

\section{Conclusion}  We have provided a map that allows one to classify suspension flows of hard particles into different regimes. This map delimits a region where dissipation is dominated by viscous forces, and where exponents describing constitutive laws and velocity fluctuations appear identical to those of frictionless particles-- these systems belong to the same universality class. In this region we believe that the floppy mode description of flows \cite{DeGiuli15a,During16} is accurate. This description is mean-field, in the sense that dissipation is assumed to occur rather homogeneously in space. 

For $\Jc \gtrsim 10^{-3}$, the leading macroscopic effect of friction is to offset $\mu_c$ and $\phi_c$, while otherwise the rheology is hardly altered \cite{Gallier14}. However, another regime, coined Frictional Sliding, can be identified sufficiently close to jamming.  In this regime, friction dominates dissipation and asymptotic behaviors differ. Dissipation appears to be localized in space, as supported by the presence of multiple velocity scales, such as the non-affine velocity and the sliding velocity. 

For realistic values of the friction coefficient we predict a transition, in an experimentally measurable regime, from a frictionless-type flow to a frictional sliding regime when the viscous number is decreased, or equivalently as the jamming point is approached.

Similar observations have been reported for inertial flows. On one hand, we showed that a similar phase diagram can be built in that case by comparing sliding and {\it collisional} dissipation \cite{DeGiuli16}. On the other hand, strong kinetic heterogeneities in frictional systems near jamming have been reported \cite{Henkes16}, corresponding in our taxonomy to the Frictional Sliding regime. Building a description of this regime and its microscopic properties characterized here remains a challenge for the future, both for inertial and over-damped flows.

Finally, the viscous and inertial limits considered here and in Ref. \cite{DeGiuli16} are two distinguished surfaces in a 3-dimensional phase diagram spanned by $\mu_p$, $J$, and the Stokes number $I^2/J$. We sketched this diagram here, and showed that transitions from viscous to inertial dynamics occur at a constant stress level, consistent with experiments \cite{Fall10}. In future work, it would be valuable to map out the intervening regions in detail.

%
%

\acknowledgments
We acknowledge discussions with B. Andreotti, L. Berthier, M. Cates, G. D\"uring and E. Lerner. M.T. acknowledges financial support by the Swedish Research Council (621-2014-4387). The simulations were performed on resources provided by the  Swedish National Infrastructure for Computing  (SNIC) at the center for scientific and technical computing at Lund University (LUNARC). M.W. thanks the Swiss National Science Foundation for support under Grant No. 200021-165509 and the Simons Collaborative Grant ``Cracking the glass problem". 

${}^\dagger$ These authors contributed equally to this work.
\bibliography{Wyartbibnew}

\end{document}